\documentclass[%
 reprint,
unsortedaddress,
 amsmath,amssymb,
 aps,
]{revtex4-1}

\usepackage{graphicx}
\usepackage{dcolumn}
\usepackage{bm}

\usepackage{siunitx}
\usepackage[version=4]{mhchem}
\usepackage[%
colorlinks,bookmarksopen,bookmarksnumbered,citecolor=blue,urlcolor=blue,anchorcolor=blue,linkcolor=blue
]{hyperref}

\begin{document}


\title{New insights into the electron trapping mechanism in \ce{LaAlO3}/\ce{SrTiO3} heterostructures}

\author{Chunhai Yin$ ^{1} $, Alexander E. M. Smink$ ^{2} $, Inge Leermakers$ ^{3} $, Lucas M. K. Tang$ ^{3} $, Nikita Lebedev$ ^{1} $, Uli Zeitler$ ^{3} $, Wilfred G. van der Wiel$ ^{2} $, Hans Hilgenkamp$ ^{2} $, and Jan Aarts$ ^{1} $}
\affiliation{
	$ ^{1} $Huygens-Kamerlingh Onnes Laboratory, Leiden University, P.O. Box 9504, 2300 RA Leiden, The Netherlands\\
	$ ^{2} $MESA+ Institute for Nanotechnology, University of Twente, P.O. Box 217, 7500 AE Enschede, The Netherlands\\
	$ ^{3} $High Field Magnet Laboratory (HFML-EMFL), Radboud University, Toernooiveld 7, 6525 ED Nijmegen, The Netherlands\\
}

\begin{abstract}
In \ce{LaAlO3}/\ce{SrTiO3} heterostructures, a commonly observed but poorly understood phenomenon is that of electron trapping in back-gating experiments. In this work, by combining magnetotransport measurements and self-consistent Schr\"odinger-Poisson calculations, we obtain an empirical relation between the amount of trapped electrons and the gate voltage. We find that the trapped electrons follow an exponentially decaying spatial distribution away from the interface. However, contrary to earlier observations, we find that the Fermi level remains well within the quantum well. The enhanced trapping of electrons induced by the gate voltage can therefore not be explained by a thermal escape mechanism. Further gate sweeping experiments strengthen our conclusion that the thermal escape mechanism is not valid. We propose a new mechanism which involves the electromigration and clustering of oxygen vacancies in \ce{SrTiO3}. Our work indicates that electron trapping is a universal phenomenon in \ce{SrTiO3}-based two-dimensional electron systems.
\end{abstract}

\maketitle

The capability of controlling the electronic properties of a material by applying an external voltage is at the heart of modern electronics. In particular in oxide heterostructures, there is a constant search to manipulate their novel functionalities with an externally applied electric field \cite{hwang2012nm}. The most well-known example is the quasi-two-dimensional electron gas (Q2DEG) discovered at the interface between \ce{LaAlO3} (LAO) and \ce{SrTiO3} (STO) \cite{Ohtomo2004}. The Q2DEG exhibits intriguing physical properties, such as superconductivity \cite{reyren2007}, signatures of magnetism \cite{brinkman2007nm,ariando2011nc,lee2013NM} and even their coexistence \cite{li2011,bert2011}. Additionally, due to the large permittivity of the STO substrate \cite{neville1972JAP}, the carrier density and mobility of the Q2DEG can be modulated by a back-gate voltage ($V\rm_{G}$). Gate-tunable insulator to metal transitions \cite{thiel2006}, insulator to superconductor transitions \cite{caviglia2008n} and Rashba spin-orbit interactions \cite{shalom2010PRL,caviglia2010PRL} have been reported. At the LAO/STO interface, the Q2DEG is confined in a quantum well (QW) on the STO side and the band structure is formed by the \ce{Ti} $t_{2g}$ orbitals. For LAO films grown on STO (001) substrates, the $d_{xy}$ band lies below the $d_{xz,yz}$ bands in energy \cite{salluzzo2009PRL,santander2011nature,smink2017PRL}. Applying $V\rm_{G}$ across the STO substrate changes the carrier density in the QW. A Lifshitz transition occurs when the Fermi level is tuned across the bottom of the $d_{xz,yz}$ bands \cite{joshua2012NC}.

In back-gating experiments, a commonly observed phenomenon is that the sheet resistance ($R\rm_{s}$) follows an irreversible route when $V\rm_{G}$ is swept first forward (increasing $V\rm_{G}$) and then backward (decreasing $V\rm_{G}$) \cite{caviglia2008n,bell2009prl,biscaras2014SR,liu2015APLM,liang2015PRB,daptary2018PRB}. The explanation as given by \citet{biscaras2014SR} is that the Fermi level lies intrinsically close to the top of the QW. High-mobility electrons escape from the QW and get trapped in STO when the carrier density is beyond a threshold. But the relations between the amount of trapped electrons, their spatial distribution, and the gate voltage are still unknown. In this Letter, we study these relations by combining magnetotransport measurements and self-consistent Schr\"odinger-Poisson calculations. We find that the earlier proposed thermal escape mechanism cannot be reconciled with our results. We perform further gate sweeping experiments which strengthen this conclusion. We propose a new mechanism which involves the electromigration and clustering of oxygen vacancies in \ce{SrTiO3}.

\begin{figure}[t]
	\centering
	\includegraphics[width=1\linewidth]{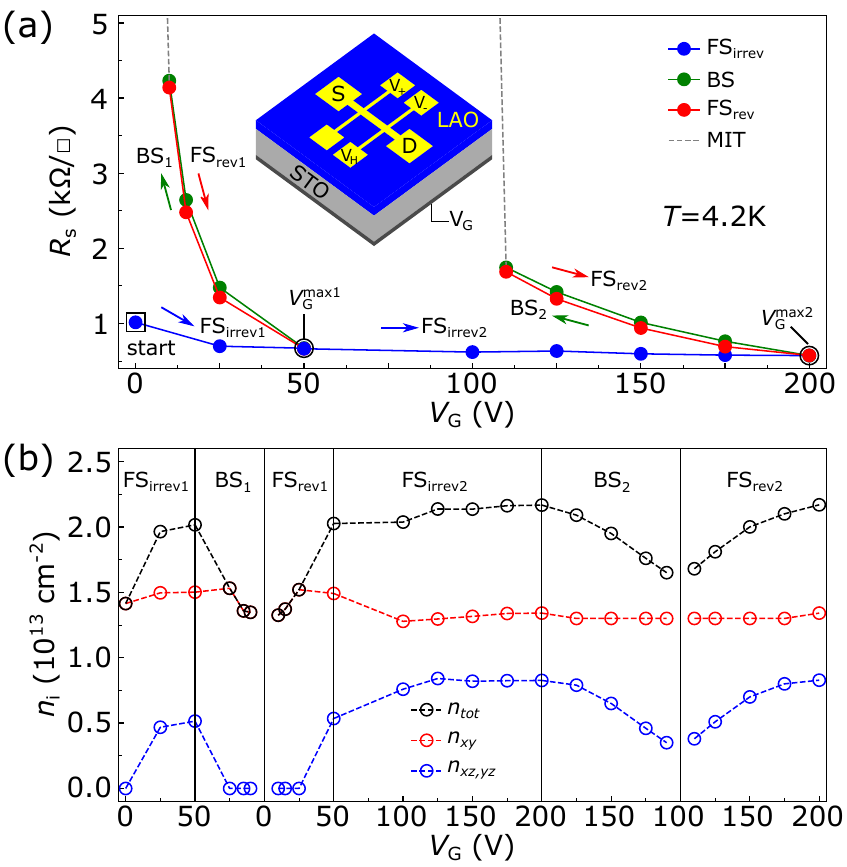}
	\caption{(a) $V\rm_{G}$ dependence of sheet resistance ($R\rm_{s}$) at \SI{4.2}{\kelvin}. The solid circles are $R\rm_{s}$($B=0$) in magnetoresistance curves. The blue, green and red arrows indicate the irreversible forward sweep (FS$\rm_{irrev}$), backward sweep (BS), and reversible forward sweep (FS$\rm_{rev}$), respectively. Two BSs were performed at \SI{50}{\volt} ($V\rm_{G}^{max1}$) and \SI{200}{\volt} ($V\rm_{G}^{max2}$). The gray dashed line indicates the metal-insulator transition (MIT). The inset shows a schematic of the Hall bar device. Source and drain are labeled as S and D. The longitudinal resistance ($R\rm_{xx}$) is measured between V$_{+}$ and V$_{-}$ and the transverse resistance ($R\rm_{xy}$) between V$\rm_{H}$ and V$_{-}$. $V\rm_{G}$ is applied between the back of the substrate and the drain. (b) $V\rm_{G}$ dependence of the carrier density in different sweeping regimes. The blue and red circles represent the carrier density of the $d_{xy}$ band ($n_{xy}$) and $d_{xz,yz}$ band ($n_{xz,yz}$). The black circle is the total carrier density ($n\rm_{tot}$) which is the sum of $n_{xy}$ and $n_{xz,yz}$.  }
	\label{fig1}
\end{figure}

The back-gating experiments were performed on Hall bar devices as depicted in the inset of Fig. \ref{fig1}(a). The length and width are \SI{1000}{\micro\meter} and \SI{150}{\micro\meter}, respectively. First, a sputtered amorphous AlO$_{x}$ hard mask in form of a negative Hall bar geometry (thickness $\sim$\SI{15}{\nano\meter}) was fabricated on a \ce{TiO2}-terminated STO (001) substrate by photolithography. Then, 15 unit cells of LAO film were deposited at \SI{800}{\degreeCelsius} in an Ar pressure of \SI{0.04}{\milli\bar} by \SI{90}{\degree} off-axis sputtering \cite{yin2018}. Finally, the sample was $in\,situ$ annealed at \SI{600}{\degreeCelsius} in \SI{1}{\milli\bar} of oxygen for \SI{1}{\hour}. The back-gate electrode was formed by uniformly applying a thin layer of silver paint (Ted Pella, Inc.) on the back of the substrate. The detailed device fabrication process is described in the Supplemental Material \cite{textsm}. The longitudinal resistance ($R\rm_{xx}$) and transverse resistance ($R\rm_{xy}$) were measured simultaneously by standard lock-in technique ($f=$ \SI{13.53}{\hertz} and $i\rm_{RMS}=$ \SI{1.0}{\micro\ampere}). Magnetotransport measurements were performed under different $V\rm_{G}$ at \SI{4.2}{\kelvin} in a superconducting magnet with the magnetic field swept between $\pm$ \SI{15}{\tesla}. The maximum applied $V\rm_{G}$ was +\SI{200}{\volt} and the leakage current was less than \SI{1}{\nano\ampere} during the measurement.

The device was cooled down to \SI{4.2}{\kelvin} with $V\rm_{G}$ grounded. Fig. \ref{fig1}(a) shows the $V\rm_{G}$ dependence of the sheet resistance, $R\rm_{s}$. $V\rm_{G}$ was first increased from \SI{0}{\volt} to \SI{50}{\volt} ($V\rm_{G}^{max1}$), resulting in a decrease of $R\rm_{s}$. This sweep is called an irreversible forward sweep (FS$\rm_{irrev}$), because $R\rm_{s}$ increased above the virgin curve when $V\rm_{G}$ was swept backward. The backward sweep (BS) led to a metal-insulator transition (MIT), which is consistent with earlier reports \cite{liu2015APLM,textvdp}. After the onset of the MIT, defined from the phase shift of the lock-in amplifier increasing above \SI{15}{\degree}, $V\rm_{G}$ was further decreased to completely deplete the QW. When $V\rm_{G}$ was swept forward again, $R\rm_{s}$ followed the same route as the BS. Therefore the latter forward sweep is named a reversible forward sweep (FS$\rm_{rev}$). Another BS was performed at \SI{200}{\volt} ($V\rm_{G}^{max2}$). It can be seen that $R\rm_{s}$ increased faster in BS$\rm_{1}$ than in BS$\rm_{2}$, and the reason of which will be discussed later. 

\begin{figure}[t]
	\centering
	\includegraphics[width=1\linewidth]{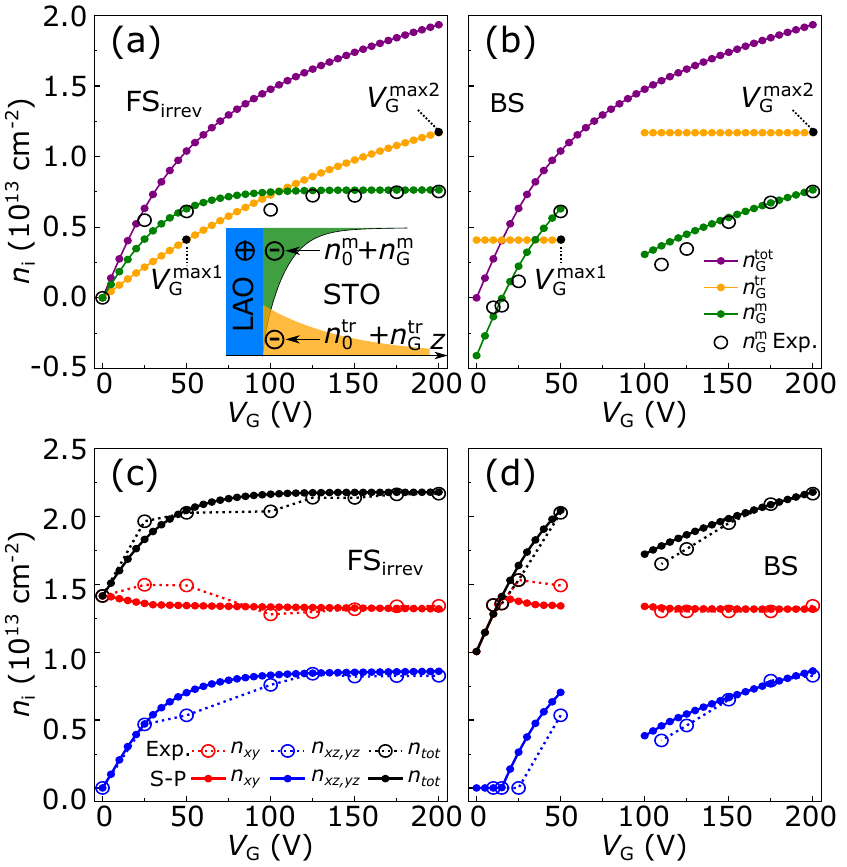}
	\caption{(a)-(b) $V\rm_{G}$ dependence of the calculated gate-induced total charge density ($n\rm_{G}^{tot}$, purple), trapped charge density ($n\rm_{G}^{tr}$, yellow), mobile charge density ($n\rm_{G}^{m}$, green) and measured gate-induced mobile charge density ($n\rm_{G}^{m}$ Exp., open black circle) in (a) FS$\rm_{irrev}$ and (b) BS regimes. The inset of (a) shows an illustration of the interface for Schr\"odinger-Poisson calculations. (c)-(d) $V\rm_{G}$ dependence of S-P calculated (solid circles) and measured (open circles) $n_{xy}$ (red), $n_{xz,yz}$ (blue) and $n\rm_{tot}$ (black) in (a) FS$\rm_{irrev}$ and (b) BS regimes.}
	\label{fig2}
\end{figure}

Fig. \ref{fig1}(b) shows the $V\rm_{G}$ dependence of the carrier density of both $d_{xy}$ ($n_{xy}$) and $d_{xz,yz}$ ($n_{xz,yz}$) bands and the total carrier density ($n\rm_{tot}$). The values were extracted by fitting the magnetotransport data with a two-band model \cite{textsm,biscaras2012PRL}. It can be seen that only the $d_{xy}$ band was occupied at \SI{0}{\volt}. In FS$\rm_{irrev1}$, electrons were added into the QW and the Lifshitz transition occurred at a carrier density ($n\rm_{L}$) of \SI{1.51e13}{\centi\meter^{-2}}, which is close to earlier reported values \cite{joshua2012NC}. In BS$_{1}$, $n\rm_{tot}$ decreased to \SI{1.33e13}{\centi\meter^{-2}} at \SI{10}{\volt}, which is the onset of the MIT, comparable to the earlier reported carrier density (0.5 -- \SI{1.5e13}{\centi\meter^{-2}}) for the MIT \cite{liao2011PRB}. In FS$\rm_{rev1}$, the carrier densities of the bands were tuned reversibly as in BS$_{1}$ and the system was fully recovered when \SI{50}{\volt} was reapplied. In FS$\rm_{irrev2}$, $n\rm_{tot}$ saturated at \SI{2.17e13}{\centi\meter^{-2}} beyond \SI{125}{\volt}. In BS$_{2}$, the MIT occurred at \SI{110}{\volt} with a carrier density of \SI{1.65e13}{\centi\meter^{-2}}, which could be due to the Hall bar contacts becoming insulating faster than the channel \cite{smink2018prb}. A noteworthy feature is that the amount of gate-induced trapped electrons is independent of the number of backward sweeps and is only related to $V\rm_{G}^\textrm{max}$.

\begin{figure*}[t]
	\centering
	\includegraphics[width=1\linewidth]{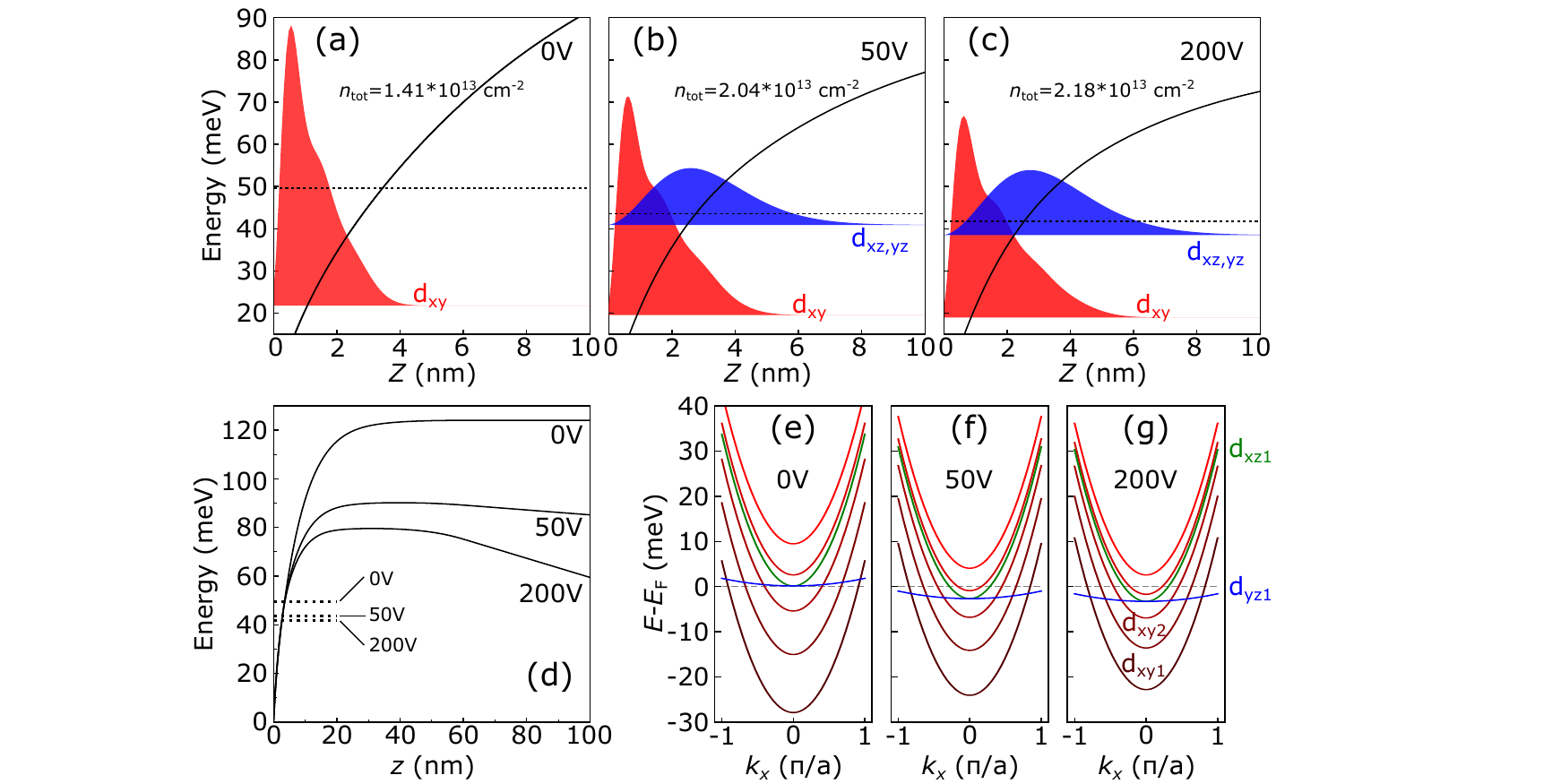}
	\caption{(a)-(c) S-P calculated confining potential profile (solid line), Fermi energy (dotted line), and spatial distribution of mobile electrons occupying $d_{xy}$ (red) and $d_{xz,yz}$ (blue) bands at (a) \SI{0}{\volt}, (b) \SI{50}{\volt} and (c) \SI{200}{\volt}. (d) S-P calculated confining potential profile and Fermi energy in a larger range. (e)-(f) S-P calculated subband dispersions in parabolic approximation at (e) \SI{0}{\volt}, (f) \SI{50}{\volt} and (g) \SI{200}{\volt}.}
	\label{fig3}
\end{figure*}

First, we study the relation between the amount of trapped electrons and the gate voltage. In a back-gating experiment, the total amount of electrons ($n\rm_{G}^{tot}$) induced by $V\rm_{G}$, as shown by the purple curves in Fig. \ref{fig2}(a) and \ref{fig2}(b), can be calculated using a parallel plate capacitor model \cite{novoselov2004Science,ihn2015}:
\begin{equation}
n{\rm_{G}^{tot}}(V{\rm_{G}}) = \frac{\epsilon{\rm_{0}}}{ed{\rm_{STO}}}\epsilon{\rm_{r}}(V{\rm_{G}})V{\rm_{G}},
\label{eqngtot}
\end{equation}
where $\epsilon{\rm_{0}}$ is the vacuum permittivity, $e$ is the electron charge and $d{\rm_{STO}}$ is the thickness of the STO substrate (\SI{0.5}{\milli\meter}). The field-dependent permittivity of the STO substrate $\epsilon{\rm_{r}}(V{\rm_{G}})$ is calculated following Ref. \cite{gariglio2015JPCM}:
\begin{equation}
\epsilon{\rm_{r}}(E) = 1+\frac{B}{[1+(E/E{\rm_{0}})^{2}]^{1/3}},
\end{equation}
where the electric field $E=V{\rm_{G}}/d{\rm_{STO}}$, $B$ = \SI{2.55e4}{} and $E{\rm_{0}}$ = \SI{8.22e4}{\volt/\meter}. In FS$\rm_{irrev}$ regimes, as shown in Fig. \ref{fig2}(a), a part of $n\rm_{G}^{tot}$ becomes gate-induced trapped electrons ($n\rm_{G}^{tr}$) in STO. Subtracting $n\rm_{G}^{tr}$ from $n\rm_{G}^{tot}$ will give the amount of gate-induced mobile electrons ($n\rm_{G}^{m}$) which are doped into the QW. We find that the relation between $n\rm_{G}^{tr}$ and $V\rm_{G}$ can be described using the following expression:
\begin{equation}
n{\rm_{G}^{tr}}(V{\rm_{G}}) = N(1-e^{-\frac{V{\rm_{G}}}{200}}),
\label{eqnt}
\end{equation}
which yields the yellow curve, where $N$ = \SI{1.85d13}{\centi\meter^{-2}}. The subtraction ($n\rm_{G}^{tot}$$-$$n\rm_{G}^{tr}$=$n\rm_{G}^{m}$) is given by the green curve, and gives a good description of the measured $n\rm_{G}^{m}$ (open black circle). In BS regimes, as shown in Fig. \ref{fig2}(b), $n\rm_{G}^{tot}$ is given by $V\rm_{G}$ according to Eq. (\ref{eqngtot}). However, the value of $n\rm_{G}^{tr}$ is fixed at the $n\rm_{G}^{tr}$($V\rm_{G}^{max}$). Thus, $n\rm_{G}^{m}$ is smaller than its counterpart in FS$\rm_{irrev}$ regimes. In both BS regimes the calculated $n\rm_{G}^{m}$ is in good agreement with the experimental data. Moreover, due to the field-dependent permittivity, $dn\rm_{G}^{tot}$/$dV\rm_{G}$ is decreasing as $V\rm_{G}$ increases. As a consequence, the same negative $\Delta$$V\rm_{G}$ removes more mobile electrons at \SI{50}{\volt} than at \SI{200}{\volt}, which could explain the fact that $R\rm_{s}$ increases faster in BS$\rm_{1}$ than in BS$\rm_{2}$. It should be noted that the empirical formula of $n\rm_{G}^{tr}$($V\rm_{G}$) is not universal, but instead varies among samples. We performed similar $V\rm_{G}$ sweeps on two reference samples and observed slightly different $V\rm_{G}$ dependence of $R\rm_{s}$ (see Fig. S5 in Ref. \cite{textsm}). Thus, $n\rm_{G}^{tr}$($V\rm_{G}$) should always be obtained from experimental results.

Next, we study the spatial distribution of the trapped electrons. The self-consistent Schr\"odinger-Poisson (S-P) model is a tool to study the charge distribution and band occupation \cite{stern1972PRB,biscaras2012PRL,scopigno2016PRL, smink2017PRL, li2018sa}. S-P calculations are based on the effective mass and envelope wave function approximations. Due to the orbital orientation, $d_{xy}$ and $d_{xz,yz}$ orbitals are heavy and light in the $z$ direction, respectively. Here, we take the effective mass as $m_{xy}^{*z}$ = 14 $m_{e}$ and $m_{xz,yz}^{*z}$ = 0.7 $m_{e}$ \cite{santander2011nature, scopigno2016PRL, smink2017PRL}, where $m_{e}$ is the electron rest mass. We take $z$ \textgreater \,0 to be STO and $z$ \textless \,0 to be LAO, as shown in the inset of Fig. \ref{fig2}(a). In the original state, there are initial mobile electrons $(n\rm_{0}^{m}$, \SI{1.41e13}{\centi\meter^{-2}} in our sample) and initial trapped electrons ($n\rm_{0}^{tr}$) on the STO side, and an equivalent amount of positive charges on the LAO side to keep overall charge neutrality.

The spatial distributions of the trapped electrons, both $n\rm_{0}^{tr}$ and $n\rm_{G}^{tr}$$(V\rm_{G})$, are input parameters of the S-P model, which effectively influence the $V\rm_{G}$ dependent occupation of the $d_{xy}$ and $d_{xz,yz}$ bands. In our calculations, we obtain the best results by using the following distribution of the trapped electrons:
\begin{equation}
n{\rm^{tr}_{3D}}(z, V{\rm_{G}}) = \begin{cases}
0 & \text{for $z < 0$} \\
\frac{n{\rm_{0}^{tr}}+n{\rm_{G}^{tr}}(V{\rm_{G}})}{\lambda}e^{-\frac{z}{\lambda}} & \text{for $z\geqslant 0$}
\end{cases} 
\label{eq3d}
\end{equation}
where $n\rm_{0}^{tr}$ = \SI{6.4e13}{\centi\meter^{-2}} and $\lambda$ = \SI{30}{\nano\meter}. The integration range is from \SI{0}{\nano\meter} to \SI{100}{\nano\meter}, which is divided into 2000 equal sections. The calculated evolutions of $n_{xy}$ and $n_{xz,yz}$ in FS$\rm_{irrev}$ and BS regimes are shown in Fig. \ref{fig2}(c) and \ref{fig2}(d), closely agreeing with the experimental data.

\begin{figure}[t]
	\centering
	\includegraphics[width=1\linewidth]{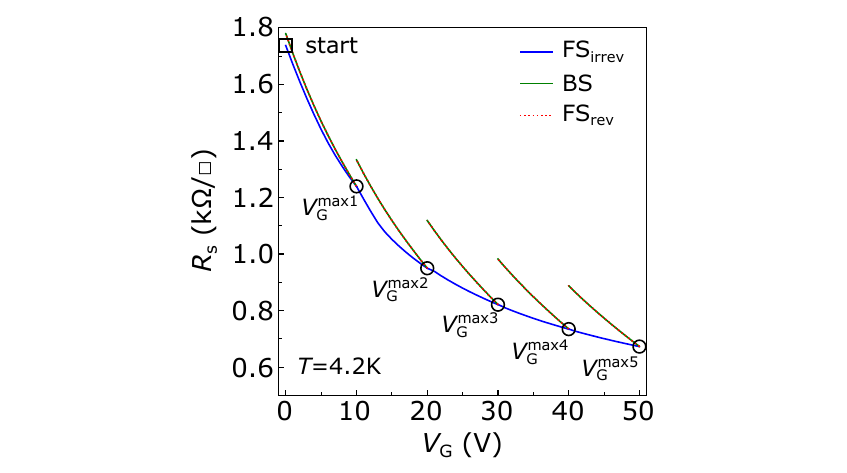}
	\caption{$V\rm_{G}$ dependence of $R\rm_{s}$ at \SI{4.2}{\kelvin}. Backward sweeps were performed from \SI{10}{\volt} to \SI{50}{\volt}. Note that BS and FS$\rm_{rev}$ overlap perfectly.}
	\label{fig4}
\end{figure}


Based on the above analysis, we could obtain the confining potential profile, the Fermi energy and the spatial distribution of mobile electrons occupying the $d_{xy}$ and $d_{xz,yz}$ bands. Fig. \ref{fig3}(a)-(c) show the results at \SI{0}{\volt}, \SI{50}{\volt} and \SI{200}{\volt}, respectively. The mobile electrons are confined within $\sim$\SI{10}{\nano\meter} at the interface, which agrees with the reported spatial distribution of the Q2DEG \cite{basletic2008nm,sing2009PRL,reyren2009apl}. Fig. \ref{fig3}(d) shows the confining potential in a larger range. It can be seen that in all cases the Fermi level is well below the top of the QW, therefore the probability of mobile electrons thermally escaping ($k_{B}T$(\SI{4.2}{\kelvin}) $\approx$ \SI{0.36}{\milli\electronvolt}) from the QW should be very low. The subband dispersions of the three cases are shown in Fig. \ref{fig3}(e)-(g). We note that increasing $V\rm_{G}$ decreases the spacing between the subband levels. 

In order to check the thermal escape mechanism \cite{biscaras2014SR} in more detail, we warmed up the device to room temperature to remove the electron trapping effect \cite{bell2009prl}. The device was cooled down to \SI{4.2}{\kelvin} again and multiple backward sweeps were performed from \SI{10}{\volt} to \SI{50}{\volt}. As shown in Fig. \ref{fig4}, a growing $R\rm_{s}$ separation between FS$\rm_{irrev}$ and BS can be clearly seen as $V\rm_{G}^\textrm{max}$ increases. In the thermal escape mechanism, it is stated that electron trapping only occurs after $R\rm_{s}$ (or $n\rm_{tot}$) reaches its saturation. However, our experiment clearly shows that electron trapping occurs immediately when positive $V\rm_{G}$ is applied and the amount of trapped electrons increases as $V\rm_{G}^\textrm{max}$ increases. So we can rule out the thermal escaping of mobile electrons to be the mechanism for electron trapping. Moreover, similar irreversible behavior has been also reported in other Q2DEG systems, such as \ce{LaTiO3}/STO \cite{biscaras2014SR}, \ce{LaVO3}/STO \cite{liang2015PRB}, (\ce{LaAlO3})$_{0.3}$(\ce{Sr2AlTaO6})$_{0.7}$/STO \cite{bal2017APL} and amorphous LAO/STO \cite{bjorlig2018APL}. Therefore the electron trapping phenomenon must be intrinsic to the STO substrate. 

\begin{figure}[t]
	\centering
	\includegraphics[width=1\linewidth]{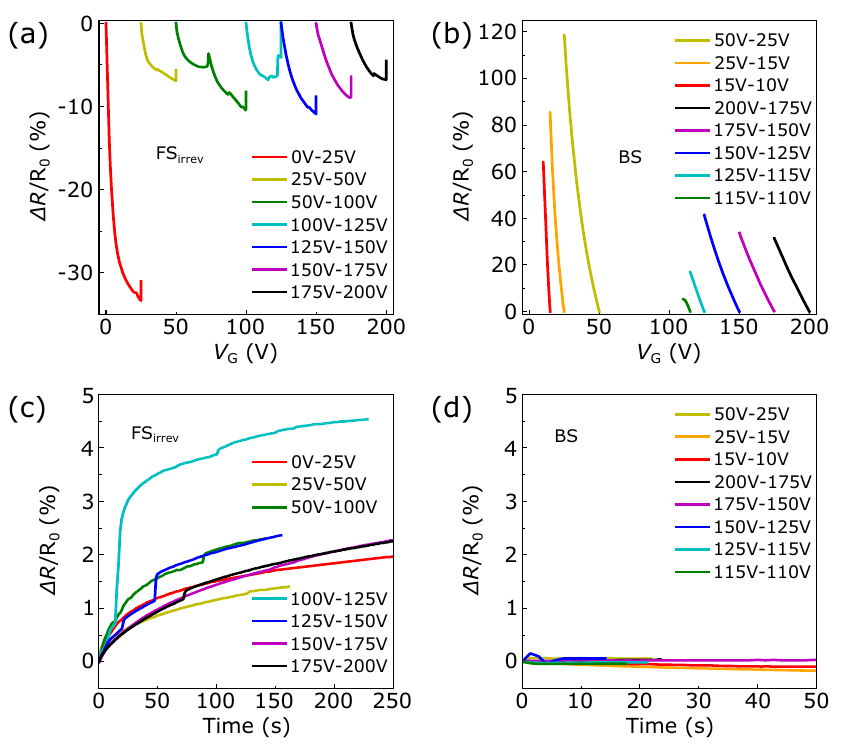}
	\caption{(a)-(b) Dynamic change of $R\rm_{s}$ during $V\rm_{G}$ sweeps in (a) FS$\rm_{irrev}$ and (b) BS regimes. $V\rm_{G}$ was swept at a rate of \SI{0.1}{\volt/\second}. $R\rm_{s}$ measurements kept on going for several minutes after the stabilization of $V\rm_{G}$. (c)-(d) Dynamic change of $R\rm_{s}$ after $V\rm_{G}$ sweeps in (c) FS$\rm_{irrev}$ and (d) BS regimes.}
	\label{fig5}
\end{figure}

We propose a two-step trapping mechanism which involves the redistribution of oxygen vacancies (V$\rm_{O}$s) in STO under influence of an electric field. The first step is the electromigration of V$\rm_{O}$s. Among all types of defects in STO, V$\rm_{O}$ has the lowest activation enthalpy for migration \cite{metlenko2014nanoscale}. Electromigration of V$\rm_{O}$s in STO has been reported in previous works \cite{hanzig2013PRB,lei2014nc,li2017AIP}. The second step is the clustering of V$\rm_{O}$s. It has been calculated that V$\rm_{O}$s clustering could form in-gap trapping states \cite{shanthi1998prb,cuong2007prl}, of which the energy was recently determined to be $\sim$\SI{0.31}{\electronvolt} and $\sim$\SI{1.11}{\electronvolt} below the conduction band \cite{baeumer2018nl}. Fig. \ref{fig5} shows the dynamic resistance change during and after $V\rm_{G}$ sweeps in the FS$\rm_{irrev}$ and BS regimes. The electron trapping mechanism can then be explained as follows. In FS$\rm_{irrev}$ regimes as shown in Fig. \ref{fig5}(a), the effect of increasing $V\rm_{G}$ is twofold. One is to add electrons into the QW. The other is to push positively charged V$\rm_{O}$s migrating toward the interface. The clustering of the accumulated V$\rm_{O}$s then forms in-gap trapping states. Several sudden resistance jumps can be clearly seen during $V\rm_{G}$ sweeps, which might be due to the formation of big V$\rm_{O}$ clusters. Moreover, after stabilizing the gate voltage as shown in Fig. \ref{fig5}(c), the electromigration and clustering of V$\rm_{O}$s do not stop immediately. Newly formed in-gap states still trap conduction electrons, resulting in an immediate increase of $R\rm_{s}$ when $V\rm_{G}$ stabilizes. In BS and FS$\rm_{rev}$ regimes as shown in Fig. \ref{fig5}(b) and \ref{fig5}(d), sweeping $V\rm_{G}$ only changes the carrier density in the QW without modifying the defect landscape near the interface. Therefore the system can be tuned in a reversible manner.

In summary, we have studied the electron trapping phenomenon in LAO/STO heterostructures under back-gate voltages. By combing magnetrotransport measurements and self-consistent Schr\"odinger-Poisson calculations, we have identified a relation between the amount of trapped electrons and the gate voltage as well as the spatial distribution of the trapped electrons. We have proposed a new trapping mechanism which involves the electromigration and clustering of oxygen vacancies in STO, since our analysis shows that the thermal escape mechanism is not valid. Our work improves the understanding of back-gating experiments in LAO/STO heterostructures. It is also valuable for theoretical works \cite{caprara2012PRL,bucheli2014PRB,scopigno2016PRL}, where the assumption was that all the gate-induced electrons land in the QW. This assumption clearly needs to be reconsidered. Equally importantly, our work indicates that electron trapping is a universal phenomenon in \ce{SrTiO3}-based two-dimensional electron systems, which is instructive to future applications of complex oxide electronic devices.

We thank Nicandro Bovenzi, Andrea Caviglia, Felix Gunkel, Kevin Steffen, Prateek Kumar and Aymen Ben Hamida for useful discussions and Anatolie Mitioglu and Lisa Rossi for their technical assistance. This work is supported by the Netherlands Organisation for Scientific Research (NWO) through the DESCO program. We acknowledge the support of HFML-RU/NWO, member of the European Magnetic Field Laboratory (EMFL). C. Y. is supported by China Scholarship Council (CSC) with grant No. 201508110214.

\newpage
{\Large Supplemental Material\\}
\vspace{0.5cm}
\section*{Hall bar device fabrication}
\begin{figure}[h]
	\centering
	\includegraphics[width=0.8\linewidth]{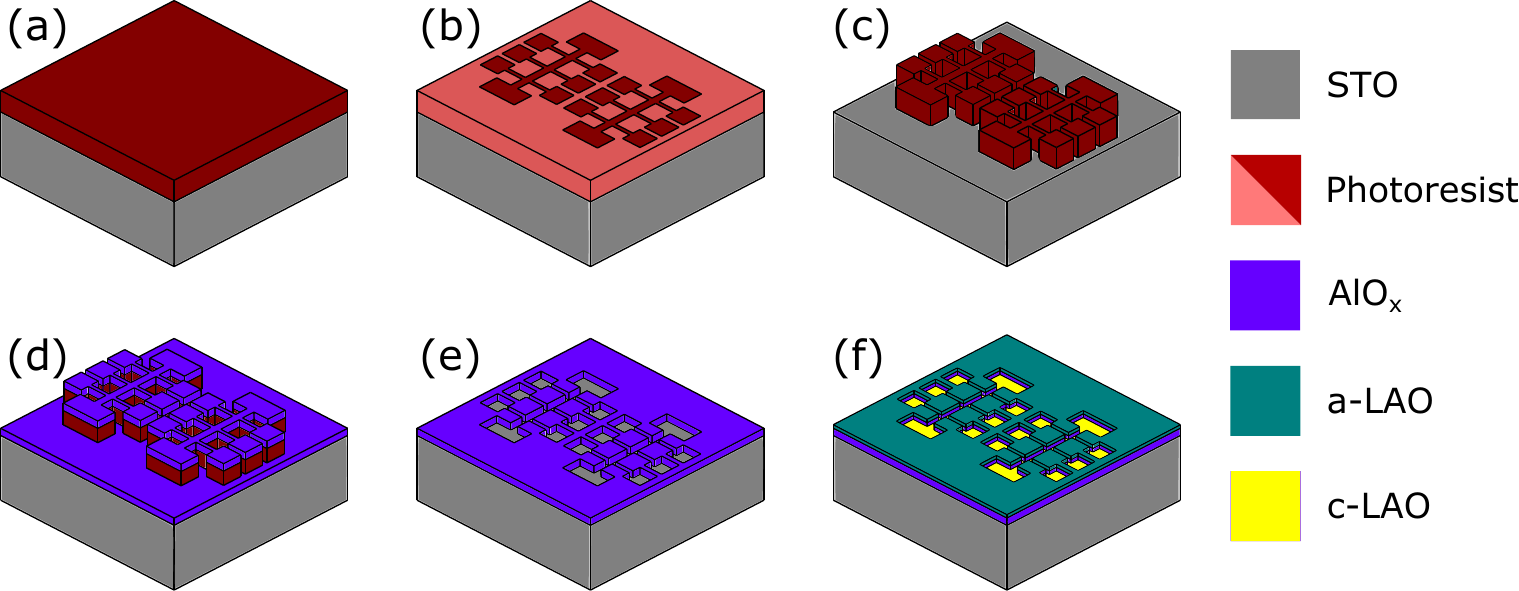}
	\caption{Schematic representation of the Hall bar device fabrication process. (a) Spin coating. (b) Exposure. (c) Development. (d) \ce{AlO_x} hard mask deposition. (e) Lift-off. (f) \ce{LaAlO3} film deposition. a-\ce{LaAlO3} and c-\ce{LaAlO3} stand for amorphous \ce{LaAlO3} and crystalline \ce{LaAlO3}, respectively. Only the interface between c-LAO and STO is conducting.}
	\label{figlith}
\end{figure}

\begin{figure}[h]
	\centering
	\includegraphics[width=0.6\linewidth]{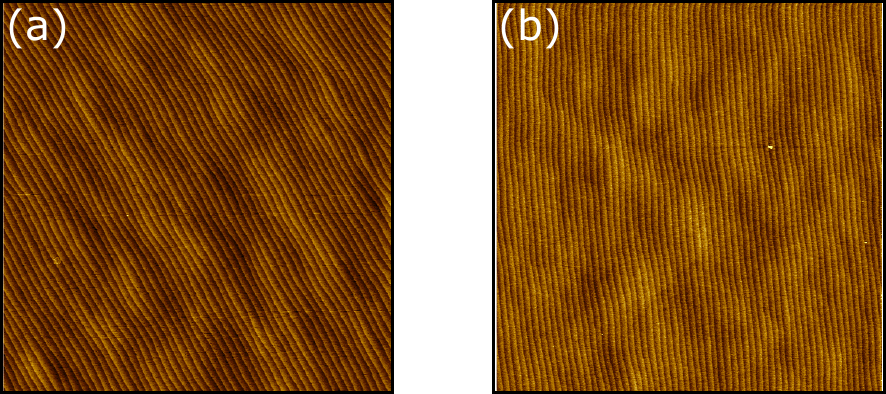}\\
	\caption{AFM images (5$\times$5 \si{\micro\meter}$ ^{2} $) of the Hall bar channel taken before (a) and  after (b) \ce{LaAlO3} film deposition.}
	\label{figafm}
\end{figure}

Before the Hall bar device fabrication, the \ce{SrTiO3} (001) substrate (5$\times$5$\times$0.5 \si{\milli\meter^{3}}) (CrysTec GmbH) was etched by buffered hydrofluoric acid and annealed at \SI{980}{\degreeCelsius} in an oxygen flow of \SI{150}{sccm} for \SI{1}{\hour} in order to form a \ce{TiO2}-terminated surface \cite{koster1998}. The device was fabricated by photolithography. First, a resist mask was patterned (Fig. \ref{figlith}(a)-(c)). The photoresist was OiR 907-12 (Fujifilm) and the exposure time was \SI{6}{\second}. The developer was OPD 4262 (Fujifilm) and the development time was \SI{1}{\minute}. Next, a \SI{15}{\nano\meter} \ce{AlO_x} layer was sputtered as a hard mask (Fig. \ref{figlith}(d)). The resist mask was then lifted by acetone (Fig. \ref{figlith}(e)). Finally, a 15 unit cells \ce{LaAlO3} film was deposited by \SI{90}{\degree} off-axis sputtering (Fig. \ref{figlith}(f)). The LAO film was grown at \SI{800}{\degreeCelsius} in an Ar pressure of \SI{0.04}{\milli\bar}, followed by an $in\,situ$ oxygen annealing treatment at \SI{600}{\degreeCelsius} in \SI{1}{\milli\bar} of oxygen for \SI{1}{\hour} \cite{yin2018}. Only the interface between crystalline LAO and STO was conducting. The width of the Hall bar was \SI{150}{\micro\meter} and the length was \SI{1.0}{\milli\meter}. AFM images of the Hall bar channel before and after LAO film deposition are shown in Fig. \ref{figafm}. Atomic steps with an average terrace width of \SI{100}{\nano\meter} can be clearly observed before and after deposition.

\newpage
\section*{Gate voltage sweeps on a film sample measured in the van der Pauw Geometry}
\begin{figure}[h]
	\centering
	\includegraphics[width=0.6\linewidth]{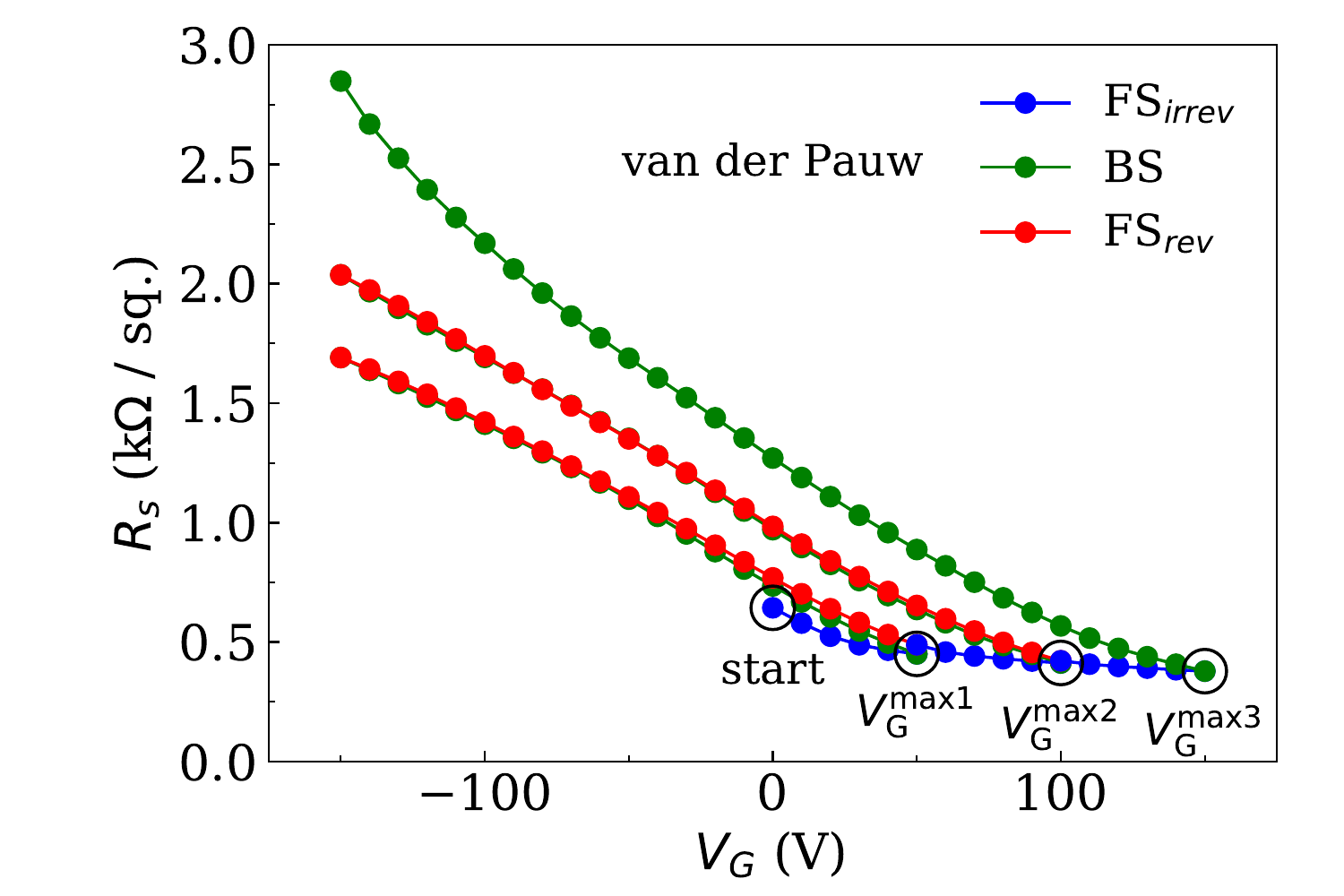}
	\caption{$V\rm_{G}$ dependence of $R\rm_{s}$ measured on an unpatterned film sample in the van der Pauw geometry.}
	\label{figvdp}
\end{figure}

We also measured an unpatterned film sample in the van der Pauw geometry at \SI{4.2}{\kelvin}. $V\rm_{G}$ was swept between $\pm$ \SI{150}{\volt}. As shown in Fig. \ref{figvdp}, the metal-insulator transition (MIT) was not observed in this case. Similar $R\rm_{s}-V\rm_{G}$ dependence has been reported in Ref. \cite{biscaras2014SR}. The explanation might be that in the van der Pauw geometry the conduction paths are not strictly defined. It is possible to measure $R\rm_{s}$ even at the maximum negative $V\rm_{G}$. While in the Hall bar device, it is easier to pinch off the narrow contacts than the wide channel, resulting in the observed MIT \cite{smink2018prb}. 

\newpage

\section*{Carrier density and mobility fitting}
\begin{figure}[h]
	\centering
	\includegraphics[width=0.45\linewidth]{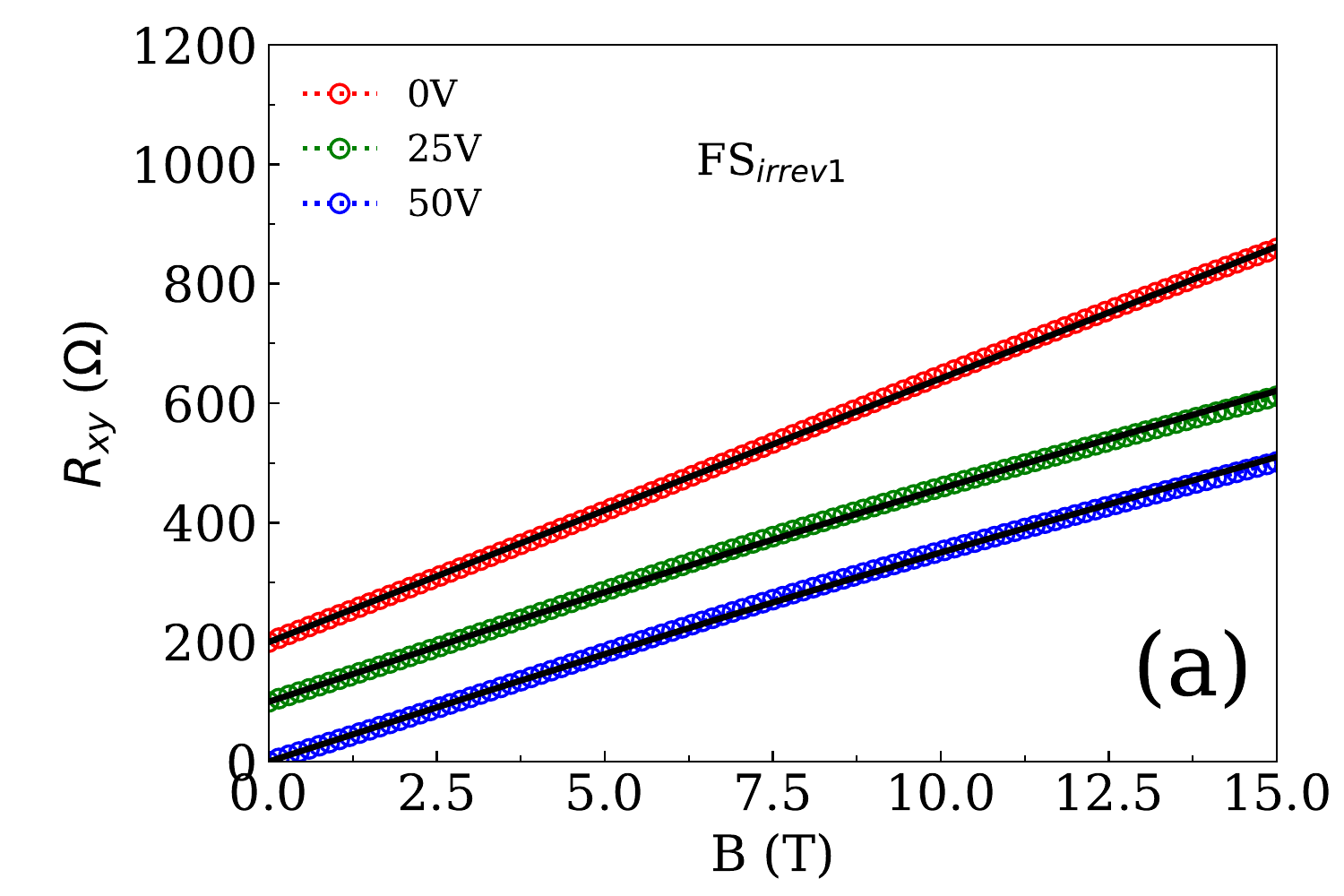}
	\includegraphics[width=0.45\linewidth]{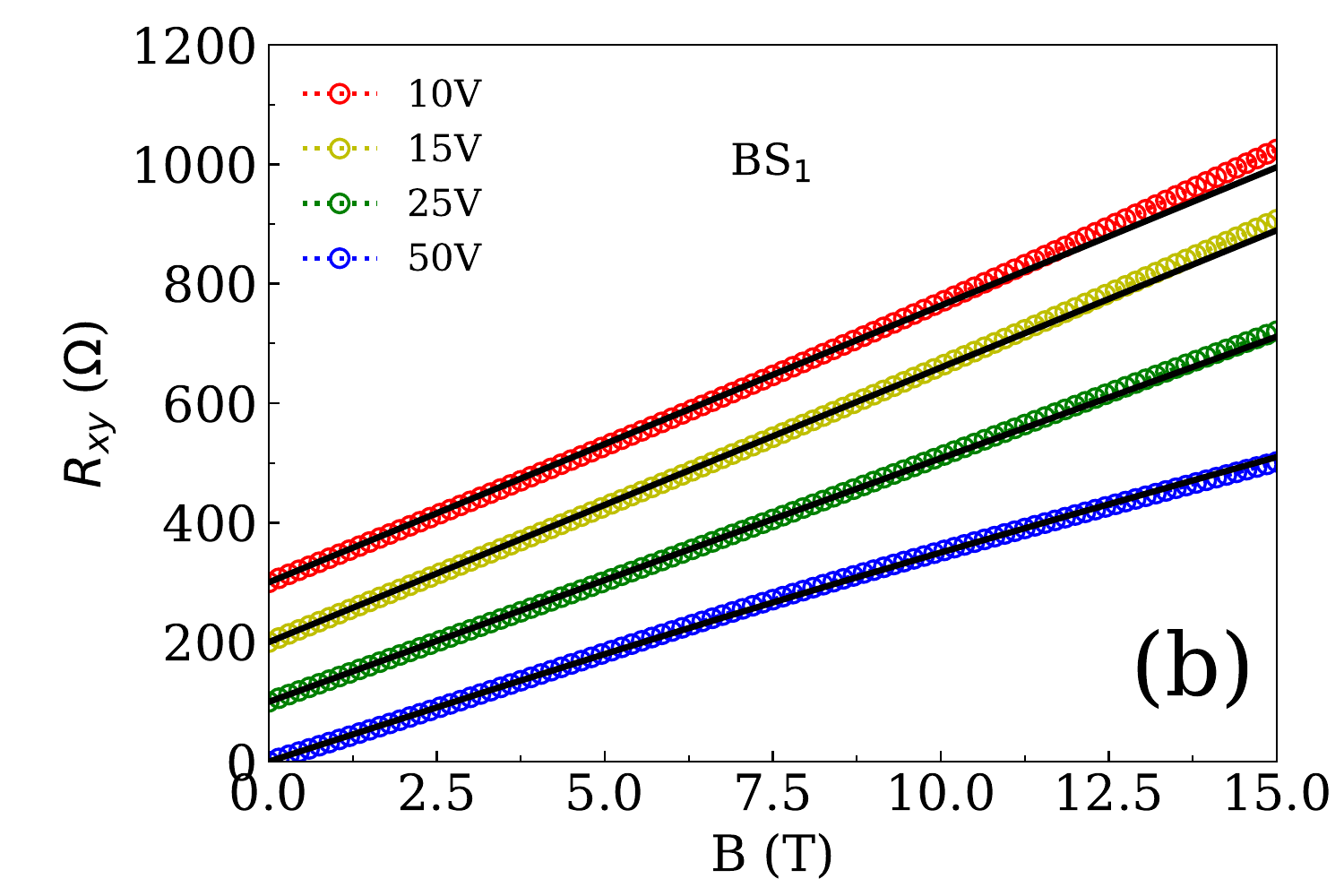}
	\includegraphics[width=0.45\linewidth]{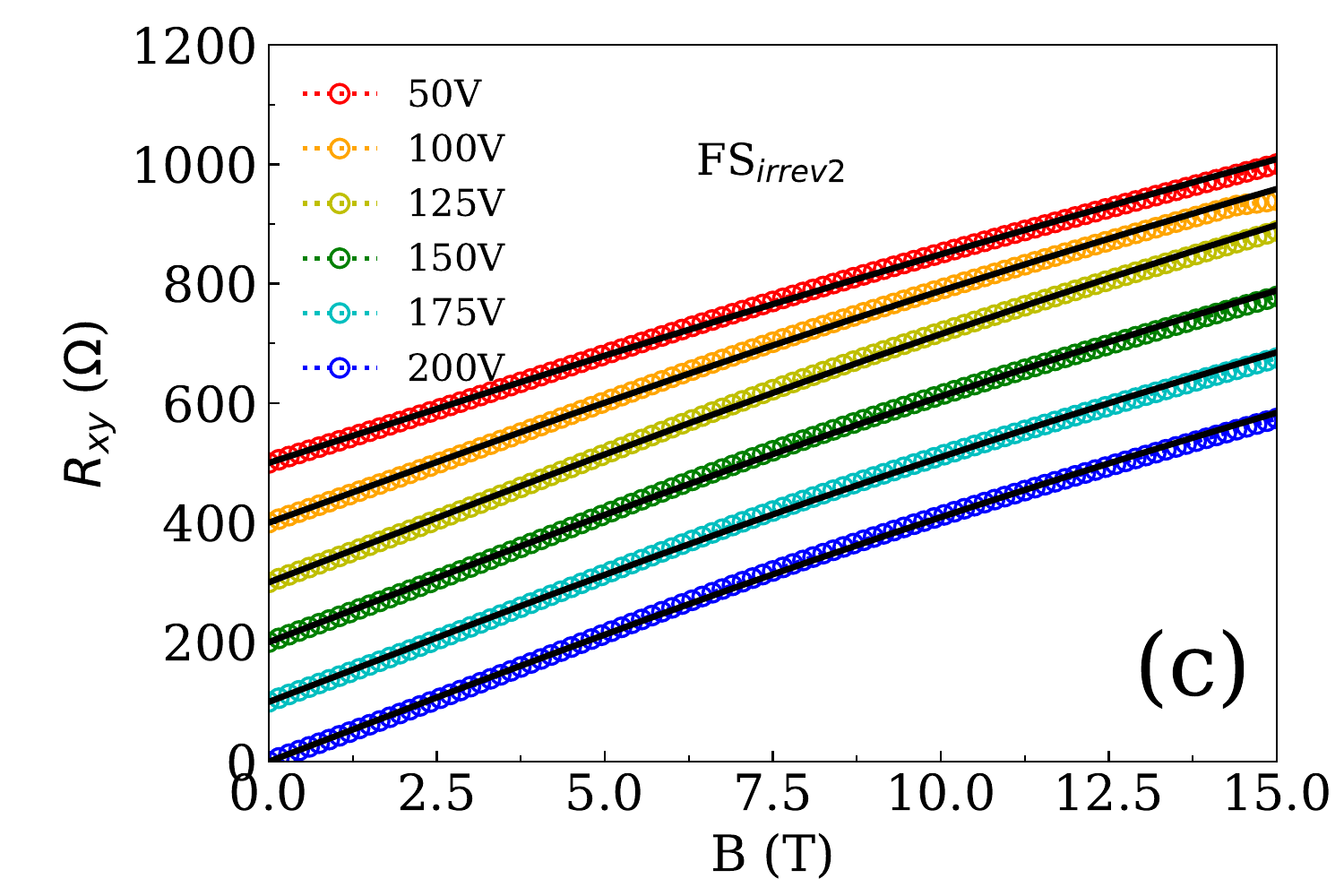}
	\includegraphics[width=0.45\linewidth]{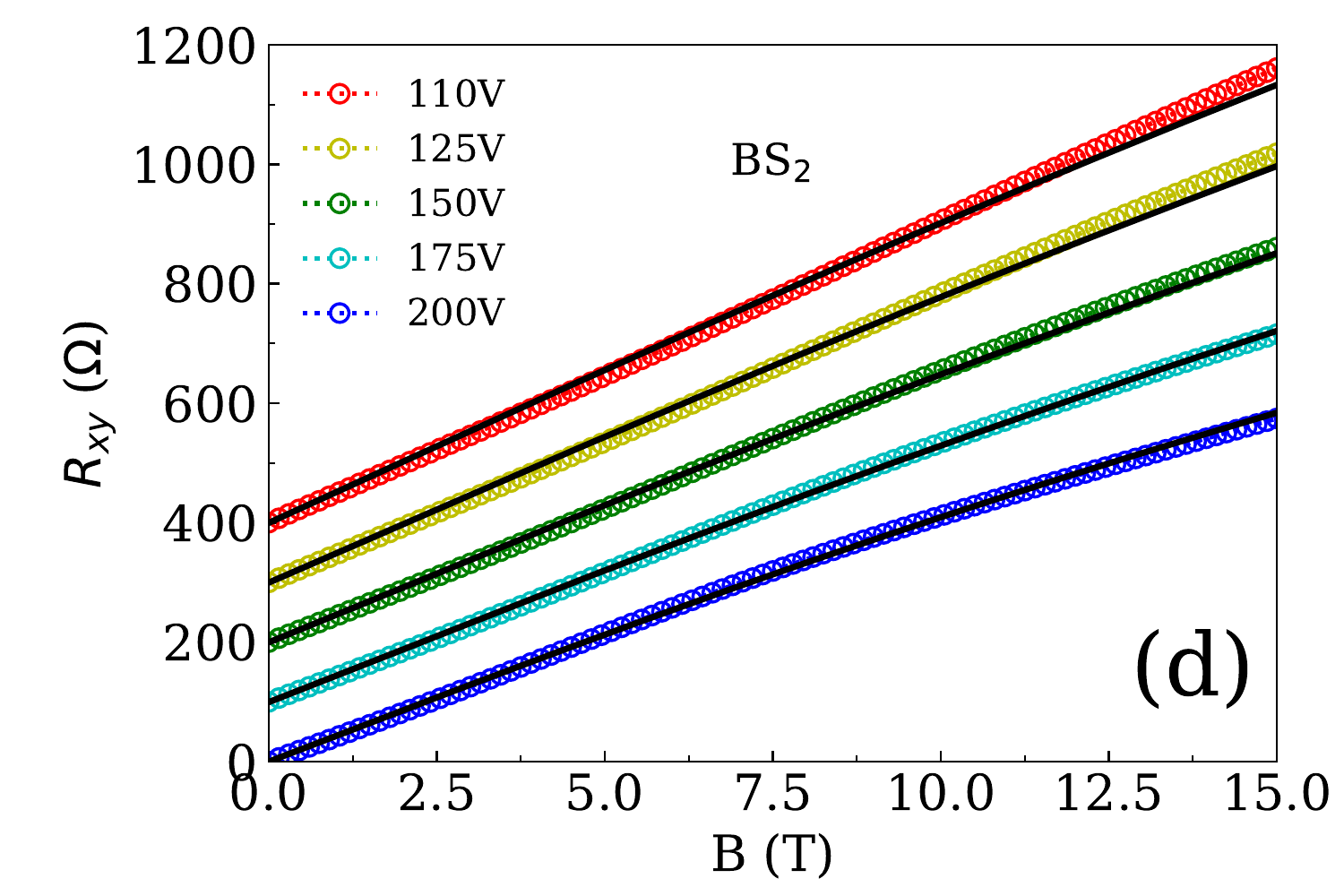}
	\caption{Two-band model fitting of the transverse resistance versus magnetic field at different gate voltages in regime of (a) First FS$\rm_{irrev}$, (b) First BS, (c) Second FS$\rm_{irrev}$, and (d) Second BS. The black lines are the fitting curves. The curves are separated by an offset for clarity.}
	\label{figfit}
\end{figure}

The carrier density and mobility of the $d_{xy}$ and $d_{xz,yz}$ bands were extracted by fitting the magnetotransport data with a two-band model \cite{biscaras2012PRL}:
\begin{equation}
R_{xy}=\frac{B}{e}\frac{\frac{n_{1}\mu_{1}^{2}}{1+\mu_{1}^{2}B^{2}}+\frac{n_{2}\mu_{2}^{2}}{1+\mu_{2}^{2}B^{2}}}{(\frac{n_{1}\mu_{1}}{1+\mu_{1}^{2}B^{2}}+\frac{n_{2}\mu_{2}}{1+\mu_{2}^{2}B^{2}})^{2}+(\frac{n_{1}\mu_{1}^{2}B}{1+\mu_{1}^{2}B^{2}}+\frac{n_{2}\mu_{2}^{2}B}{1+\mu_{2}^{2}B^{2}})^{2}},
\end{equation}
with the constraint $1/R\rm_{s}=en_{1}\mu_{1}+en_{2}\mu_{2}$, where $n_{1}$ and $n_{2}$ are the carrier densities of the $d_{xy}$ and $d_{xz,yz}$ bands, and $\mu_{1}$ and $\mu_{2}$ are the corresponding mobilities. The fitting of the irreversible forward sweep (FS$\rm_{irrev}$) and backward sweep (BS) regimes is shown in Fig. \ref{figfit}. The fitting of the reversible forward sweep (FS$\rm_{rev}$) regimes is similar to that of BS, which is therefore omitted.

\newpage

\section*{Gate voltage sweeps on reference devices}
\begin{figure}[h]
	\centering
	\includegraphics[width=0.6\linewidth]{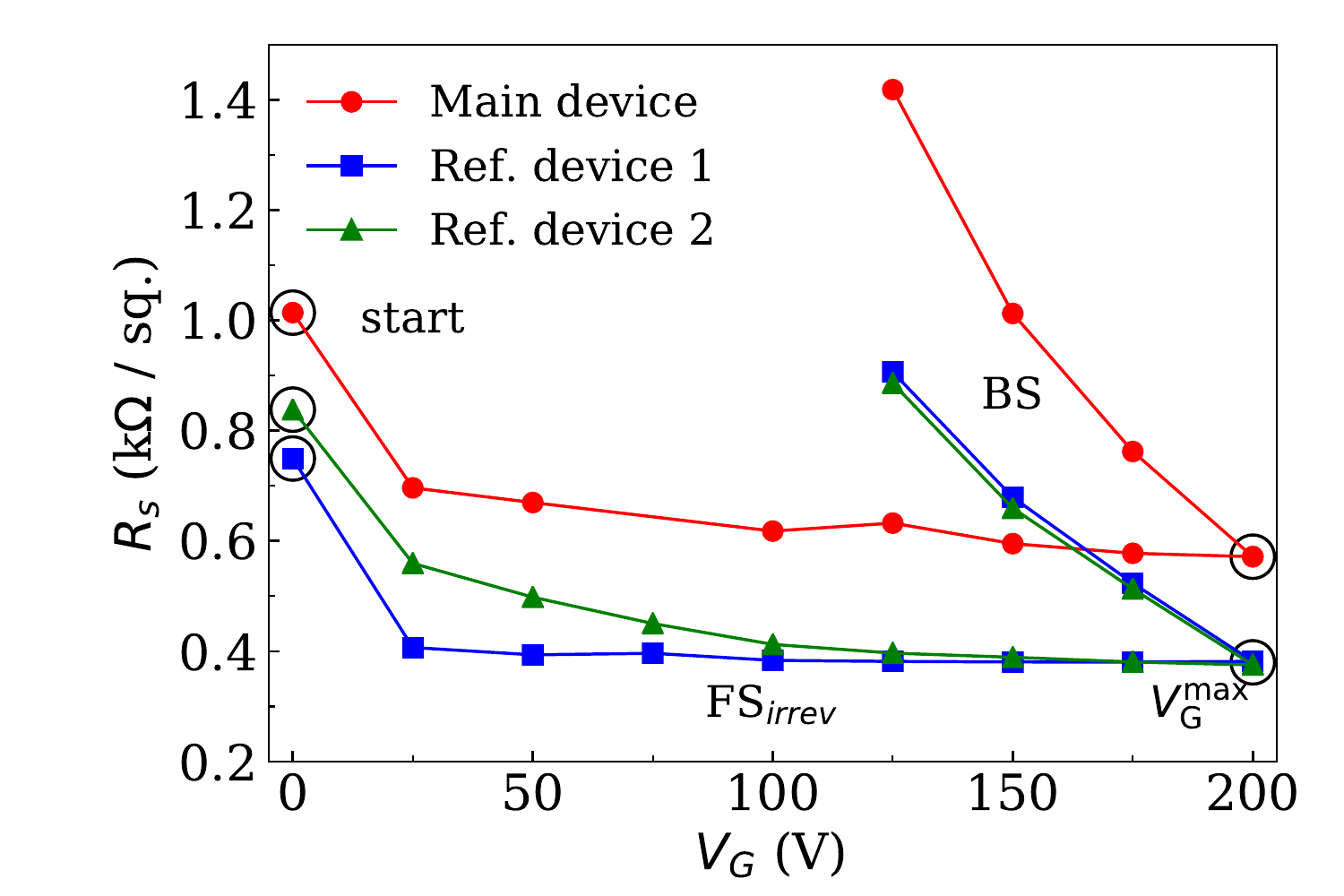}
	\caption{$R\rm_{s}$ versus $V\rm_{G}$ of two reference devices at \SI{4.2}{\kelvin}.}
	\label{figref}
\end{figure}

Two reference devices were fabricated under the same conditions as described above and the same $V\rm_{G}$ sweeps were performed. As shown in Fig. \ref{figref}, the two samples behave slightly differently, which indicates that the amount of defects and their distribution at the interface are different from sample to sample. 

\newpage

\bibliographystyle{apsrev4-1}
\bibliography{LAO2}

\begin{thebibliography}{49}
\bibitem [{\citenamefont {{It should be noted that the MIT can only be observed
  in Hall bar devices}}()}]{textvdp}%
  \BibitemOpen
  \bibfield  {author} {\bibinfo {author} {\bibnamefont {{It should be noted
  that the MIT can only be observed in Hall bar devices}}},\ }\href@noop {}
  {}\bibinfo {howpublished} {but not in unpatterned film samples measured by
  the van der Pauw method. See Fig. S3 in Ref. [22]}\BibitemShut {NoStop}%
\end{thebibliography}

\end{document}